\documentclass[letterpaper, 10 pt, conference]{ieeeconf}  

\IEEEoverridecommandlockouts                              

\usepackage{graphicx}
\usepackage{caption}
\usepackage{url}
\usepackage{hyperref}

\usepackage{algorithm}

\usepackage[commentColor=blue]{algpseudocodex}
\algrenewcommand\textproc{} 

\usepackage{subcaption}  
\usepackage{hhline}
\usepackage{multirow}
\usepackage{placeins}
\usepackage{amssymb}

\renewcommand{\baselinestretch}{0.97}
\usepackage[modulo,switch]{lineno} 

\usepackage{amsmath}

\title{\LARGE \bf
Conflict-Based Search as a Protocol: A Multi-Agent Motion Planning Protocol for Heterogeneous Agents, Solvers, and Independent Tasks
}

\author{
Rishi Veerapaneni$^{1}$, Alvin Tang$^{\dagger1}$, Haodong He$^{\dagger2}$, Sophia Zhao$^{\dagger1}$, Viraj Shah$^{\dagger1}$, Yidai Cen$^{\dagger1}$, Ziteng Ji$^{\dagger3}$, \\
Gabriel Olin$^{\ddagger1}$, Jon Arrizabalaga$^{\ddagger1}$, Yorai Shaoul$^{\ddagger1}$, Jiaoyang Li$^{1}$, Maxim Likhachev$^{1}$
\thanks{This work was partially supported by National Science Foundation (NSF) grant \#2328671, by NSF Graduate Research Fellowship Program grant \#DGE2140739, and a gift from Amazon. This research partially used Bridges-2 at PSC \cite{brown2021Bridges2} through the ACCESS program which is supported by NSF grants \#2138259, \#2138286, \#2138307, \#2137603, and \#2138296. } 
\thanks{$\dagger, \ddagger$ Equal authorship respectively, sorted alphabetically}
\thanks{$^1$ Carnegie Mellon University, USA}
\thanks{$^2$ Tongji University, China}
\thanks{$^3$ UC Berkeley, USA}
\thanks{Corresponding author: vrishi@cmu.edu}
\thanks{Webpage: \url{https://rishi-v.github.io/CBS-Protocol/} }
}

\begin{document}

\maketitle
\thispagestyle{empty}
\pagestyle{empty}

\begin{abstract}

Imagine the future construction site, hospital, or office with dozens of robots bought from different manufacturers. How can we enable these different robots to effectively move in a shared environment, given that each robot may have its own independent motion planning system?
This work shows how we can get efficient collision-free movements between algorithmically heterogeneous agents by using Conflict-Based Search (Sharon et al. 2015) as a protocol.
At its core, the CBS Protocol requires one specific single-agent motion planning API; finding a collision-free path that satisfies certain space-time constraints. Given such an API, CBS uses a central planner to find collision-free paths - independent of how the API is implemented.
We demonstrate how this protocol enables multi-agent motion planning for a heterogeneous team of agents completing independent tasks with a variety of single-agent planners including: Heuristic Search (e.g., A*), Sampling Based Search (e.g., RRT), Optimization (e.g., Direct Collocation), Diffusion, and Reinforcement Learning.

\end{abstract}



\section{Introduction}
The rapid progress of robotics technology is lowering robots' costs while enhancing capabilities, leading to a future in which multi-agent robotic systems will be ubiquitous.
Such systems may operate across a wide range of domains, including construction, manufacturing, office environments, and even sophisticated household settings. Importantly, these multi-agent teams are unlikely to be homogeneous; rather, they will consist of robots produced by different manufacturers, each performing independent tasks. This raises a fundamental question: how can heterogeneous robots, each potentially equipped with its own motion planning system, coordinate their movements in a shared environment?

In particular, such robots may:
\begin{enumerate}
\item Exhibit different embodiments and kinodynamic constraints,
\item Pursue distinct and independent tasks, and
\item Be developed by different companies, each with proprietary motion planning solvers.
\end{enumerate}

While this may seem far-fetched, it is already beginning to appear in practice. Current automated warehouses and manufacturing facilities often integrate diverse platforms such as forklifts, trucks, and mobile carts, which must all operate within the same workspace. However, it remains unclear how to enable such heterogeneous robots to navigate and coordinate effectively in a common environment.



To that end, our main contribution is demonstrating how we can use Conflict-Based Search as a protocol to enable efficient collision-free movement. 
Conflict-Based Search (CBS) \cite{sharon2015cbs} is a foundational method from Multi-Agent Path Finding \cite{stern2019mapfbenchmark} that finds collision-free multi-agent paths.
At its core, we view CBS as defining a protocol that requires one specific single-agent motion planning API; finding a path that avoids space-time locations. Given such an API, CBS uses a central planner to resolve collision between agents - independent of how the API is implemented. Thus, ``CBS as a protocol" enables collision-free multi-agent motion planning using different single-agent planners. We demonstrate the effectiveness of this protocol by showing how we can effectively plan for a heterogeneous team with a variety of single-agent planners including: Heuristic Search, Sampling Based Search, Optimization, Diffusion, and Reinforcement Learning. Additionally, the CBS Protocol enables each agent to complete different independent tasks (e.g., coverage, surveillance) instead of just start-goal tasks.

\subsection{Contributions}
Taking a step back, to our knowledge the problem of Multi-Agent Motion Planning (MAMP) with algorithmically heterogeneous solvers has not been explored by the MAMP community (discussed more in Section \ref{sec:ah-mamp}). Thus, our first contribution is to bring attention to this problem which we term Algorithmically Heterogeneous MAMP (AH-MAMP).

Second, the CBS Protocol bridges a gap between single-agent motion planning, multi-agent motion planning, and Multi-Agent Path Finding (MAPF). In particular, our CBS Protocol solution takes existing technology (i.e., existing single-agent solvers and existing MAPF techniques) and shows how they can be put together in a novel way.
Thus, our paper is a proof-of-concept paper whose primary contributions are shaped by the reader's context:


(1) Readers focused on single-agent motion planning but not MAMP; For these readers, the key contribution of this paper is showing how we can use their single-agent motion planners in multi-agent systems through the CBS Protocol via defining one single API. The hope is that the protocol more easily enables single-agent motion planning researchers to incorporate their planners into multi-agent systems. 

(2) Readers working on real-world MAMP but are unfamiliar with CBS, or only familiar with CBS in its classical (gridworld) context; For these readers, the primary contribution is showing how the CBS Protocol extends beyond gridworld and can incorporate a range of diverse motion planners (e.g., optimization) for kinodynamic motion planning.

(3) Readers familiar with CBS and its extensions past non-gridworld; For these readers, the main contribution is the protocol formulation using heterogeneous solvers at once as well as some intricacies of the single-agent planners in the CBS context (e.g., using an RL policy). We note that the CBS Protocol high-level search does not differ from CBS.


\section{Related Work}
Our work focuses on collision-free multi-agent motion planning (MAMP) for agents with independent tasks.

We first describe what we call ``algorithmically homogeneous" MAMP methods that plan for agents using the same method across all agents, even if the agents themselves are heterogeneous. These methods explicitly require access to the internals of the agents and typically extend a single-agent planning technique to multiple agents. We then describe ``algorithmically heterogeneous" MAMPF methods that plan for agents which have different single-agent planners.



\subsection{Algorithmically Homogeneous MAMP Methods} \label{sec:alg-homo-mamp}
We describe a range of algorithmically homogeneous MAMP methods based on their broader taxonomy. 


\subsubsection{Search-Based Planning} 
Search-Based Planning techniques for MAMP typically focus on a simplified problem set-up called Multi-Agent Path Finding (MAPF) \cite{stern2019mapfbenchmark}. In the ``classical" MAPF set-up, the world is discretized into a graph where each vertex corresponds to a grid cell. Time is also discretized into timestep where each agent can move to an adjacent cell or wait at its current location at each timestep (these define edges between the corresponding vertices). Each agent occupies exactly one vertex/grid cell at a timestep. Search-based MAPF solvers use variants of A* space-time solvers.
Operator Decomposition \cite{standley2010operater_decomposition} addresses the combinatorial increase in search size by incrementally generating successors by planning individual agents. M* \cite{mstart_2011} uses a similar approach to plan for up to 100 2D agents. 

Conflict-Based Search \cite{sharon2015cbs} is a foundational technique in MAPF. We describe CBS more in-depth in Section \ref{sec:cbs-explanation}. Conceptually, CBS is a two-level heuristic search method that iteratively detects and resolves conflicts. CBS utilizes a high-level search that searches over constraints and a low-level search the plans single-agent paths. CBS first plans each agent individually and detects conflicts (i.e., collisions). CBS then proceeds to iteratively resolve conflicts by adding constraints to specific agents and replanning until a collision-free solution is found. 
Although originally created for MAPF with single-agent A* solvers, CBS has been used with other single-agent solvers as we will see later in this section. Several methods use different variants of A* solvers to handle heterogeneous teams of car-like robots with kinematic constraints \cite{wen2022carlike_cbs,bai2025cbs_heterogeneous_robots} or manipulator arms \cite{shaoul2024generalized_ecbs}.


\subsubsection{Sampling-Based Planning} 
Multi-Agent RRT* \cite{vcap2013multi_agent_rrt} is an initial attempt that tries to plan for up to 10 2D agents by running RRT* on the combined state space. 
Multi-Robot Discrete RRT (MRdRRT) \cite{solovey2016discrete_rrt} is able to plan for up to a combined 60 degrees of freedom over various agents with different degrees of freedom. MRdRRT's main insight is to plan over a discrete implicit graph and to modify the extend operation to move agents sequentially in a priority order.

\cite{solis2021representation_optimal_cbs} plans for a heterogeneous team by generating a probabilistic roadmap for each agent and using CBS.
Safe Interval Continuous-space CBS \cite{sim2024safe_interval_rrt} recently proposed a novel single-agent Safe Interval RRT* with CBS to plan for up to 100 2D agents in continuous space.

\subsubsection{Optimization Methods}
Optimal Reciprocal Collision Avoidance (ORCA) \cite{alonso2013orca} is able to plan for 1000's of 2D agents by having each agent independently solve a low-dimensional linear program based on the positions and velocities of nearby agents. \cite{wang2017safety_barrier_certificates} introduces using safety control barrier certificates via formulating and solving a quadratic programming problem. 
Recently introduced Space-Time Graphs of Convex Sets \cite{tang2025multiagent_space_time_gcs} utilizes agent priorities and convex optimization to find solutions for up to 10 2D agents.

\cite{debord2018multi_drone_ground_cbs} first uses discrete planning (A*) in CBS and then post-processes the solution with optimization to plan for a heterogeneous drone and ground robot team.
Discontinuity-Bound Conflict-Based Search (db-CBS) \cite{moldagalieva2024dbcbs} interleaves discrete search and optimization by using a discontinuity-bounded A* search (e.g., A* with motion primitives) in CBS to generate single-agent approximate solutions which are used as warm-starts for a joint optimization across all agents. If the optimization does not return a feasible solution, then constraints are added and the CBS search continues. 
Conflict-Based Model Predictive Control (CB-MPC) \cite{tajbakhsh2024cbmpc} replaces the low-level A* search in CBS with limited horizon MPC.

\subsubsection{Machine Learning Methods}
Multi-Agent DDPG \cite{lowe2017maddpg} and Multi-Agent PPO \cite{yu2022mappo} extrapolate single-agent DDPG \cite{lillicrap2015ddpg} and PPO \cite{schulman2017ppo} respectively for multi-agent systems. These approaches are more general than collision-free MAMP but can be used for MAMP (e.g., \cite{peng2023mappohr}).

Several methods train a single agent policy for MAPF using reinforcement learning or supervise learning that learns to avoid collisions 
\cite{sartoretti2019primal} 
and can handle up to 10000 agents in congestion \cite{jiang2024sillm}. 
Other recent work MMD \cite{shaoul2025mmd} trained a diffusion model as a single-agent planner for CBS and incorporates constraints through guidance terms.

The main restriction with all these algorithmically homogeneous MAMP methods is that they require all agents to run the same planner. Thus, these methods cannot incorporate agents manufactured / programmed externally which limits their application in such instances. The CBS Protocol directly addresses this limitation.

Taking a step back, we note that CBS appears in each category with different types of single-agent solvers, with their emphasis on the single-agent solver. CBS as a Protocol has a different focus and can be seen as encompassing all these methods. Our perspective of CBS using an abstracted single-agent API means that we can recover \cite{sharon2015cbs,wen2022carlike_cbs,bai2025cbs_heterogeneous_robots,shaoul2024generalized_ecbs,solis2021representation_optimal_cbs,sim2024safe_interval_rrt,debord2018multi_drone_ground_cbs,tajbakhsh2024cbmpc,shaoul2025mmd} with different API implementations.



\subsection{Algorithmically Heterogeneous MAMP Techniques} \label{sec:ah-mamp}
Unlike algorithmically homogeneous MAMP methods, algorithmically heterogeneous MAMP methods do not require each agent to run the same solver. 
To our surprise, we could not find any published work that addresses this problem setting. Existing MAMP methods for heterogeneous teams focus on robots with different capabilities but use algorithmically homogeneous solutions (e.g., \cite{bai2025cbs_heterogeneous_robots,solis2021representation_optimal_cbs,debord2018multi_drone_ground_cbs}). On the other hand, existing multi-agent task planning/coordination methods focus on heterogeneous behaviors or task assignment and not on collision-free movement \cite{rizk2019heterogeneous_multi_robot_survey,bettini2023heterogeneous_behavior_marl}. 



Thus, part of this paper's goal is to introduce / bring attention to the Algorithmically Heterogeneous MAMP (AH-MAMP) problem setting. AH-MAMP tries to achieve collision-free motion planning for heterogeneous single-agent solvers without being able to modify the solvers. Solutions for AH-MAMP instead require designing multi-agent \textit{protocols} with well-defined single-agent APIs, with the protocol/API abstraction enabling using heterogeneous single-agent solvers.

To our knowledge, current solutions for algorithmically heterogeneous multi-robot teams in the real world currently plan robots individually viewing other robots as dynamic obstacles. The most common solution is for agents to plan independent paths and rely on local sensing to replan when another agent obstructs their path. However, this reduces their ability to move fast, requires additional planning time during execution, and can lead to deadlock.
An alternative scheme could be to use Prioritized Planning \cite{erdmann1987multiple} which assigns agents priorities and plans agents sequentially starting with the highest priority agent. Each agent plans a path that avoids colliding with the paths of previously planned (higher priority) agents. However, in congestion this can result in no solution being found, which limits its practicality.
Taking a step back, the problem with both of these approaches is that they solve the multi-agent problem as greedy single-agent calls viewing other agents as obstacles, resulting in them failing in complex instances. 
To that end, we introduce CBS as a Protocol as a principled AH-MAMP protocol that does not view agents as obstacles but instead intelligently searches over space-time constraints to find collision-free solutions.

\begin{figure*}[t!]
    \centering
    \includegraphics[width=0.90\linewidth]{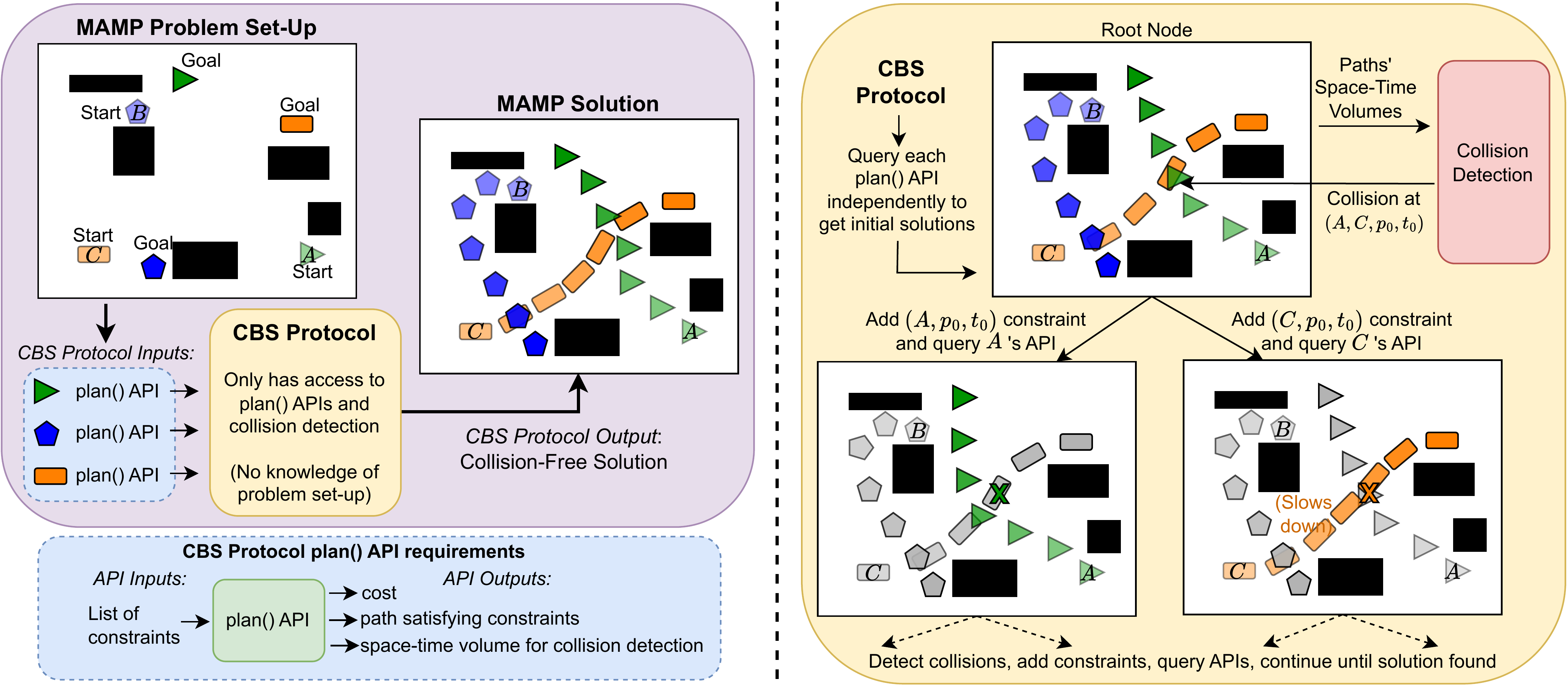}
    \caption{We depict using the CBS Protocol (left) and its internal process (right). Left: Given a Multi-Agent Motion Planning (MAMP) problem, each agent defines a \texttt{plan()} API that satisfies the CBS Protocol requirements (blue dashed box). This is input into the protocol (which has no other knowledge except for collision detection) which uses the APIs and outputs a collision-free solution. Right: The CBS Protocol starts with generating a root node by querying each agent's plan() API without constraints. It detects collisions between agents (e.g., agents A and C collide spatially at point $p_0$ at time $t_0$) and generates two child CT nodes. Each node is created by adding a constraint (colored X) that the agent needs to avoid $(p_0,t_0)$ and re-querying the agent's plan() API with the additional constraint. CBS repeats this process of detecting collisions, adding constraints, and querying APIs with additional constraints until a collision-free solution is found.}
    \label{fig:cbs-protocol}
    \vspace{-1em}
\end{figure*}

\section{CBS as a Protocol}
We first describe CBS from classical MAPF and then describe CBS as a protocol.

\subsection{Conflict Based Search} \label{sec:cbs-explanation}
CBS is a two-level heuristic search algorithm for solving MAPF optimally. We focus on the top-level which searches over ``Constraint Tree" (CT) nodes and describe its core search components (e.g., nodes, goal condition, etc).
Algorithm \ref{alg:alg-cbs-protocol} describes the CBS Protocol (identical to CBS except for the \texttt{plan} API) and can be used to follow along.

\textbf{CT Nodes:}
At the high level, CBS employs a best-first search over CT nodes. Each CT node contains 1) a set of constraints, and 2) a set of agent paths, one for each agent, that satisfy the constraints. Each constraint prevents an agent from occupying certain space-time locations and is used to avoid collisions with other agents. 
Conceptually, each CT node represents a candidate solution.

\textbf{Goal Condition:}
A CT node is considered a goal node if its agents paths have no collisions (line \ref{line:goal-condition}).

\begin{algorithm}[t]
\caption{CBS Protocol}
\label{alg:alg-cbs-protocol}
\textbf{Parameters}: Agents $a_i$ with \texttt{plan()} API \\ \noindent
\textbf{Output}: Collision free paths
\begin{algorithmic}[1] 

\Procedure{CBSProtocol}{$\{a_i\}$}
    \State node = generateRoot() \label{line:root-node}
    \State OPEN.insert(node)
    \While{OPEN.notEmpty() and not TimeOut()}
        \State node $\gets$ OPEN.pop() \label{line:OPEN-pop}
        \State conflicts $\gets$ detectConflicts(node.volumes) \label{line:detect-conflicts}
        \If{conflicts $= \emptyset$} \Comment{Found solution} \label{line:goal-condition}
            \State \textbf{return} node.paths
        \EndIf
        \State constraints $\gets$ conflicts[0].getConstraints() \Comment{Resolve the first conflict via constraints} \label{line:get-constraints}
        \For{$(a_i, p, t) \in$ constraints} \Comment{Simplified} \label{line:going-through-constraints}
            \State n $\gets$ node.copy() \Comment{New node} \label{line:creating-successor}
            \State n.cons[$a_i$].append($(p,t)$)
            \State response $\gets$ $a_i$.\texttt{plan}(n.cons[$a_i$]) \label{line:api-query}
            \If{response is None}
                \State \textbf{continue} \Comment{Skip if API did not find solution}
            \EndIf
            \State n.paths[$a_i$], n.volumes[$a_i$], n.costs[$a_i$] $\gets$ response
            \State OPEN.insert(n) \label{line:OPEN-insert}
        \EndFor
    \EndWhile
    \State \textbf{return} No solution found
\EndProcedure
\end{algorithmic}
\end{algorithm}

\textbf{Successor Generator:}
Suppose we have a CT node with a set of agent paths and constraints. If the paths have no collisions, then this is a goal node and CBS terminates. However, if it does have a collision, then CBS chooses one conflict and resolves it by generating two child CT nodes (lines \ref{line:get-constraints}-\ref{line:OPEN-insert}) with \textit{mutually disjunctive} constraints (a pair of constraints such that a collision occurs if both are violated). 

For example, suppose two agents collide at a specific vertex $v_0$ at timestep $t_0$, denoted $(v_0,t_0)$. A valid solution cannot have both of those agents at $(v_0,t_0)$, so CBS generates two possible candidate solutions; one with the first agent prevented from visiting $(v_0,t_0)$ and the second with the second agent prevented from visiting $(v_0,t_0)$.
This results in two children CT nodes (one for each possibility). Importantly, each children node differs from the parent node by just having an additional constraint on one agent. Thus, generating a children CT node just requires replanning that one agent while keeping the other agents' paths the same.
The replanning is typically done by a ``low-level" heuristic search algorithm (e.g., A*).

\textbf{Initial ``Root" CT Node:}
CBS starts by generating the ``Root" CT node which contains no constraints (line \ref{line:root-node}). Generating the Root CT node requires generating paths for each agent individually (this is the only node that requires planning all agents). Once generated, this node is inserted into the CT queue and the high-level CT search is started.

\textbf{High-Level Search:}
CBS employs a best-first search over the CT nodes by using a priority queue (which is called ``OPEN"). There are multiple ways to sort the priority queue based on the MAPF objective, but one standard way is to sort by CT node's cost (typically the sum of each agent's path length). CBS pops out the cheapest node and checks if it is a goal node (line \ref{line:OPEN-pop}). If it is, CBS returns that node's paths as the solution. If not, CBS generates successor nodes and inserts them into OPEN and repeats (line \ref{line:OPEN-insert}).

\textbf{Summary:}
CBS searches over candidate solutions (high-level search over CT nodes) by iteratively finding collisions, adding constraints, and replanning individual agents (low-level search). CBS is a centralized planner that plans full-horizon collision-free paths for all agents.

\textbf{Properties: }
In the classical MAPF form with optimal high-level A* search and optimal low-level A* planners, CBS is complete and optimal (i.e., it will find the optimal solution given enough time and memory).

\subsection{CBS as a Protocol}
As mentioned in Sec \ref{sec:alg-homo-mamp}, although CBS has predominantly used A* single-agent planners, the general idea of iteratively planning and resolving collisions by searching over a Constraint Tree can be applied to other single-agent planners \cite{moldagalieva2024dbcbs,sim2024safe_interval_rrt,tajbakhsh2024cbmpc,shaoul2025mmd}. However, those publications with CBS focus on developing the single-agent planner.


We come from the AH-MAMP perspective where we do not focus on single-agent solvers but instead want a mechanism to coordinate an algorithmically diverse set of agents.
Our insight is that CBS can be used as a protocol across multiple different solvers at once. CBS solely requires each agent to have the ability to plan a path that satisfies the CT node's constraints. Thus, each agent could run their own individual solver with their own kinematic constraints. Since different agents could have different cost functions (other than path length), each solver should also return the solution cost, which can be included in the CT node.

An important caveat is that given a set of agent plans, CBS must detect collisions between agents. 
Thus, we require the single-agent planner to not just return a path but also the space-time volume occupied by it. Note that compact representations for this are possible (e.g., if the robot's footprint is static, giving the footprint along with the space-time path defines the occupied space-time volume).


CBS as a protocol is exactly the same as CBS except for the change of perspective of the single-agent planner. In particular, instead of the planner being predefined, we view the plan call as an API call that can be written independently.

\textbf{Single-Agent Solver \texttt{plan()} API:}
\begin{itemize}
    \item Input: List of constraints
    \item Output: 1) Path that satisfies the constraints, 2) Space-time volume occupied by the path (likely through a compact representation), 3) Solution cost \\
    Or None if no such path found
\end{itemize}

The CBS protocol can now be used to plan collision-free paths independently of how the single-agent API is defined, enabling using different solvers. 
Additionally since we only require the output of the solver, this allows using proprietary solvers that do not want to share their internal workings.

In the context of CBS, this API will be repeatedly queried with different constraints. If the API is solving the same task between calls (e.g., start-goal task), this means the API is repeatedly solving the same problem with minor variations (through the constraints). \cite{shaoul2024experience-cbs} showed how CBS can be sped up by having its single-agent solver reuse ``experience" from previous queries. This idea is broadly applicable to different types of single-agent motion planning solvers. 
Thus single-agent solver APIs are recommended to leverage experience from past queries for faster query times in the CBS Protocol.

\textbf{Conflict Detection and Constraints:} 
The CBS Protocol detects inter-agent collisions in a CT node's solution by checking for overlaps between all pairs of agents' space-time volume. If a collision is detected between two agents $A, C$, a representative space-time point $(p_0,t_0)$ in the overlapped region is selected (in 2D this corresponds to a point $p_0 =(x_0,y_0)$). To resolve this collision, we add a $(\text{agent\_id},p_0,t_0)$ constraint for each colliding agent that forbids the agent from overlapping spatially with $(p_0,t_0)$ (we note prior work \cite{li2019largeAgentMAPF} introduced this constraint for large agents in gridworld MAPF).
We highlight that the constraint is defined in the \textit{workspace} dimension (e.g., 2D or 3D) and is independent of the dimensionality of each agent / solver.

A subtle but important caveat is that in continuous time motion planning, a constraint at a specific instantaneous time could be satisfied by planning a near identical path with an almost instantaneous wait, which will result in another collision at the same location at a negligibly offset time. Likewise for continuous space motion planning, a constraint at a single point could be satisfied by planning a near identical path with an almost negligible deviation.
Thus, for practical purposes, the CBS Protocol should use constrain space-time volumes (e.g., forcing the agent to avoid $p_0$ from $(t_0, t_0+\delta)$ where $\delta$ is the constraint duration) that force meaningful changes in agents' paths. 
We note that there are extensive improvements in the CBS literature for designing better constraints \cite{shaoul2024generalized_ecbs,srli2021,andreychuk2021improving}, but this is not related to our main contribution and should be incorporated in the future. 

\textbf{Independent Tasks:} CBS for motion planning has been primarily used to plan collision-free start-goal paths. From the API perspective, however, the API simply needs to find a solution that satisfies the constraints. A few prior works have used this flexibility in CBS for planning paths that visit multiple goal locations \cite{ren2022cbs_steiner}. Thus more generally, the API is not limited to planning start-goal paths and can be used for arbitrary single-agent tasks (e.g., coverage, exploration).

We highlight that the CBS Protocol requires independent tasks between agents. If agents need to explicitly collaborate (e.g., move an object together) while avoiding collisions with other agents, our protocol is not applicable for two reasons. First, collaborative behavior between two agents is robot / task specific while the CBS Protocol is robot / task agnostic. Second, collaborative behavior is fundamentally not reasoned about in CBS, which only focuses on resolving collisions. 

If there are different groups of collaborative agent that need to work together, then the CBS Protocol could be used via meta-agents. Users could specify a single API that plans for all agents in a group (this is a ``meta-agent"), which can then be input into the CBS Protocol to ensure that groups do not collide with each-other. Internally, each group API would run their own custom collaborative planning scheme.

\textbf{Theoretical Properties of the CBS Protocol: }
The theoretical guarantees of using the CBS Protocol depend on the single-agent APIs and the constraints used. 
The CBS Protocol is ``complete" (i.e., guaranteed to eventually find a valid solution in a solvable instance) given two conditions:
1) Each single-agent API will find a valid solution with bounded cost that satisfies constraints if a solution exists, and 
2) The constraints are mutually disjunctive.
The proof of completeness follows regular CBS \cite{sharon2015cbs}.

However, as mentioned earlier, for practical performance constraints should be on space-time volumes, which are not mutually disjunctive. More broadly, in continuous time planning, designing meaningful mutually disjunctive constraints for CBS is an unsolved question \cite{li2025cbscontinuoustimerevisit}.
Thus, the CBS Protocol is not theoretically complete in the continuous space-time context, however, it still offers a principled way to resolve inter-agent conflicts for efficient motion planning.

\section{Experiments}
Our experiments seek to show the generality of the CBS Protocol by using a variety of solvers from different single-agent motion planning families. Our experiments focus on motion planning in a 2D occupancy map with agents with different kinematic constraints, solvers, and tasks. 

\subsection{Single-Agent Solvers}
Table \ref{tab:single-agent-solvers} shows a summary of the different solvers which we implemented the required \texttt{plan()} single-agent API for the CBS Protocol. We briefly describe each solver here and how they incorporate space-time constraints and experience.

\begin{table*}[t]
    \centering
    \begin{tabular}{|c||c|c|c|c|} \hline
    Solvers & Space-Time Constraints & Experience & Dynamic Constraints & Tasks \\ \hline
    A* & Edge validity & Successor Generator \cite{shaoul2024experience-cbs} & Ackermann, 4-connected, Dubins & Start-goal, Coverage \\
    RRT & Edge validity & Re-use partial valid tree & 4-connected, None & Start-goal, Coverage \\
    Direct Collocation & Constraint & Warm start & Ackermann & Start-goal, Surveillance \\
    Diffusion & Penalty & Warm start \cite{shaoul2025mmd} & None & Start-goal, Motion patterns \\
    RL & Action masking & Caching & Unicycle & Start-goal, Coverage \\ \hline
    \end{tabular}
    \caption{Various Single-Agent Solvers used in the experiments with CBS-Protocol. We note that these solvers cover different motion-planning paradigms (heuristic search, sampling, optimization, diffusion, RL respectively).}
    \label{tab:single-agent-solvers}
    \vspace{-1em}
\end{table*}

\subsubsection{Heuristic Search}
We implement a single-agent A* search as a representative heuristic search algorithm. 
We plan in continuous space for robots with dynamic constraints (i.e., Ackermann and Dubins) by using discretized motion primitives. We use a modified version of soft-duplicate detection \cite{du2019soft_duplicate_detection} to speed up search.

\textit{Constraints:} Incorporating space-time constraints is straightforward; we search over space-time nodes and prune edges/nodes that violate space-time constraints.

\textit{Experience:} We implement experience by using the technique in \cite{shaoul2024experience-cbs}. When we find a solution, we store the nodes on the solution path. Then when replanning, if we encounter a node in the solution path, we add as successors the rest of the nodes in the path that are valid under the new constraints. 


\subsubsection{Sampling Based Search}
We implement Rapidly-exploring Random Tree (RRT) \cite{lavalle1998rrt} as a representative of sampling based search algorithms. 
We use a regular spatial RRT for its simplicity and popularity (with each vertex having an associated time of arrival). We note this means that the RRT solution does not contain wait actions and that the solver is incomplete. However, it can still be used in a plan() API (it may just return failure when solutions exist).


\textit{Constraints:} 
Space-time constraints are incorporated in the edge/node validity function.


\textit{Experience:} 
Given a new query, we iterate through all nodes/edges in the tree created in the previous query and remove those invalid under the new constraints. We then start the RRT process from this tree.
We note more sophisticated versions of replanning using experience exist (e.g., \cite{ferguson2006replanning_with_rrts}).

\subsubsection{Optimization}
We formulate the motion planning problem as a trajectory optimization problem, whose objective is to drive the system from the initial state to the goal in minimum time. To this end, we transcribe the problem into a nonlinear program, with a cost function that minimizes a weighted objective and constraints that enforce both dynamic feasibility and spatial validity. Dynamic constraints are implemented using direct collocation~\cite{kelly2017introduction}, while collision avoidance is addressed through a duality-based exact representation~\cite{zhang2020optimization}. We initialize the nonconvex nonlinear program with an A* solution and solve it using IPOPT~\cite{wachter2006implementation}.

\textit{Constraints:} 
We handle space–time constraints explicitly, in the same way as dynamics and spatial constraints.

\textit{Experience:} We incorporate experience by warm starting the optimization with the previous solution. 

\subsubsection{Diffusion}
We implement a diffusion solver by modifying Multi-robot Multi-model planning Diffusion (MMD) \cite{shaoul2025mmd} for our map and tasks.

\textit{Constraints:} MMD incorporates space-time constraints by adding a guidance term into the diffusion process that penalizes the distance between the agents path and the $(x,y)$ location of an active constraint.

\textit{Experience:} MMD generates paths by initializing a path with complete random noise and denoising it. MMD uses experience by replacing the random noise initialization with the path of a previous solution with some slight noise added.

\subsubsection{Reinforcement Learning}
We implement an Reinforcement Learning (RL) solver using PPO \cite{schulman2017ppo}. To our knowledge, we are unaware of existing papers that have used an RL single-agent solver with CBS.
The model takes in a local observation and outputs a categorical probability distribution of the next action. Concretely, the model inputs are:
\begin{enumerate}
    \item An estimate of distance to goal computed via a backward Dijkstra assuming gridworld dynamics (so unaware of the agent's footprint or dynamics),
    \item Local obstacle distances by ray casting at 30 degree increments (e.g., simulating a lidar scan), and
    \item Previous action indices
\end{enumerate}
The model outputs a 1-hot probability distribution corresponding to a discrete set of motion primitives for a robot with Dubins dynamics. We utilize motion primitives to handle kinodynamic constraints and as it also enables using action masking to speed up training.
To generate a full path, we rollout the policy, greedily following its actions and backtracking if necessary when hitting a deadend (i.e., DFSing based on the agent's policy).

\textit{Constraints:} 
We incorporate space-time constraints during test time by masking actions that violate constraints.
We also tried training a constraint aware version (i.e., constraints were also input into the model) but that performed worse. 


\textit{Experience:} 
We cache the previous solution and given a new call, reuse the previous solution up to the earliest violated constraint, minus a small slack (5 simultation seconds).



\begin{figure}[t!]
    \centering
    \includegraphics[width=0.95\linewidth]{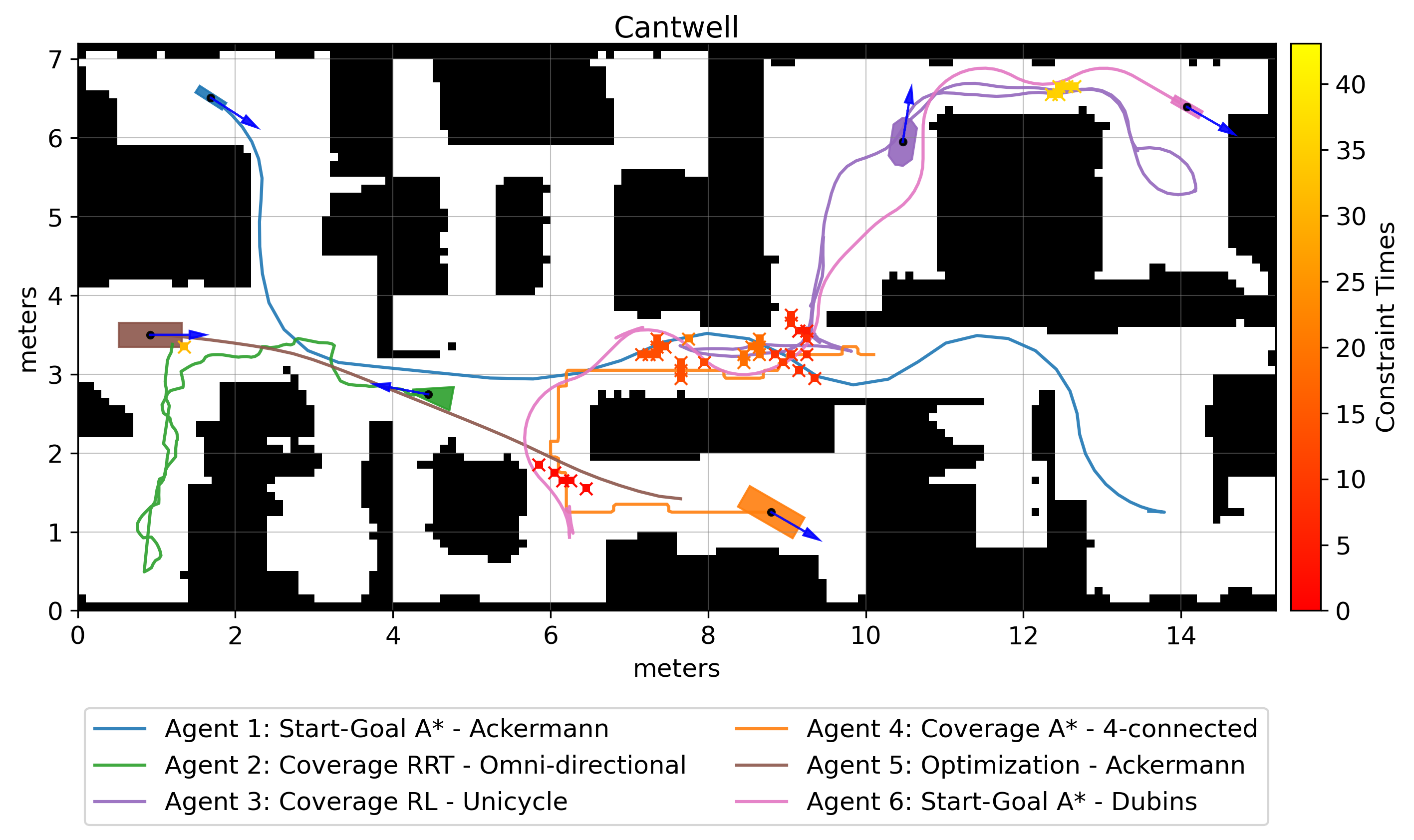}
    \caption{An example of the CBS Protocol with 6 heterogeneous agents with different solvers, dynamics, and tasks. Constraints for the solution are plotted as colored x's.}
    \label{fig:case-study}
    \vspace{-1em}
\end{figure}

\subsection{CBS Protocol Settings}
We primarily care about the success rate (as opposed to solution cost) in our experiments and therefore use a ``Greedy" CBS variant \cite{barer2014suboptimalecbs} for the CBS Protocol that prioritizes CT nodes by number of pairs of agents with conflicts and tiebreaks by solution cost. We use discrete collision checking every 0.1 simulation seconds to detect conflicts. Each robot has a fixed footprint, so the APIs output space-time volumes by outputting their space-time path and footprint, which is then used by CBS for collision detection. Constraints for a collision at $(x_0,y_0,t_0)$ block a spatial square of size 0.1 containing $x_0,y_0$ for 2.5 seconds. These numbers were semi-arbitrarily picked for performance and to emphasize continuous space / continuous time constraints and planning.

\subsection{Experimental Results} \label{sec:main-results}
We test our method on the Cantwell (7.2x15.2, Fig \ref{fig:case-study}), Ribera (8.7x6.3), and Mosquito (23.4x11.1) maps from the IN2D dataset \cite{dobrevskiskocaj2020in2d} that realistically model a house or office space, as well as an empty map (10x15) and large (32x32) map with 10\% obstacles (size in meters). 
Each robot has a footprint between 0.1 to 0.8 meters (some circles, rectangles, and triangles). All single-agent APIs/solvers, collision-checking, and the CBS Protocol were written in Python and experiments were run on a machine with an i7-11800H@2.30GHz x 16 CPU and 3050 Ti Laptop GPU.

\textbf{Start-Goal Planning:}
We evaluate the CBS Protocol's ability to handle a variety of different solvers at once. Each solver was given a per-query time limit of 5-10 seconds with an overall CBS Protocol time limit of 120 seconds. The A* and RRT solvers worked on most instances on all the maps, while RL and Collocation struggled on the IN2D maps, and Diffusion worked well only on the Ribera map. For each multi-agent problem instance with $N$ agents, we semi-randomly select $N$ start-goal pairs, and then sample a random solver that is able to solve that pair. This means that the distribution of solvers and maps in our experiments is not uniform. We generated instances with 2-10 agents through this mechanism which resulted in 675 instances.

Recall from Section \ref{sec:cbs-explanation} that the initial ``root node" created in CBS requires planning each agent individually without constraints. Thus, the number of ``root conflicts" (number of pairs of agents that conflicts) serves as an approximation for problem difficulty as having more conflicts implies a more congested problem instance that requires more work for the CBS Protocol to resolve.
Additionally, a root node with 0 conflicts indicates that individually planning and executing would result in a successful instance. Of the 675 instances, 175 instances fell into this category.

\begin{figure}[t!]
    \centering
     \includegraphics[width=0.99\linewidth]{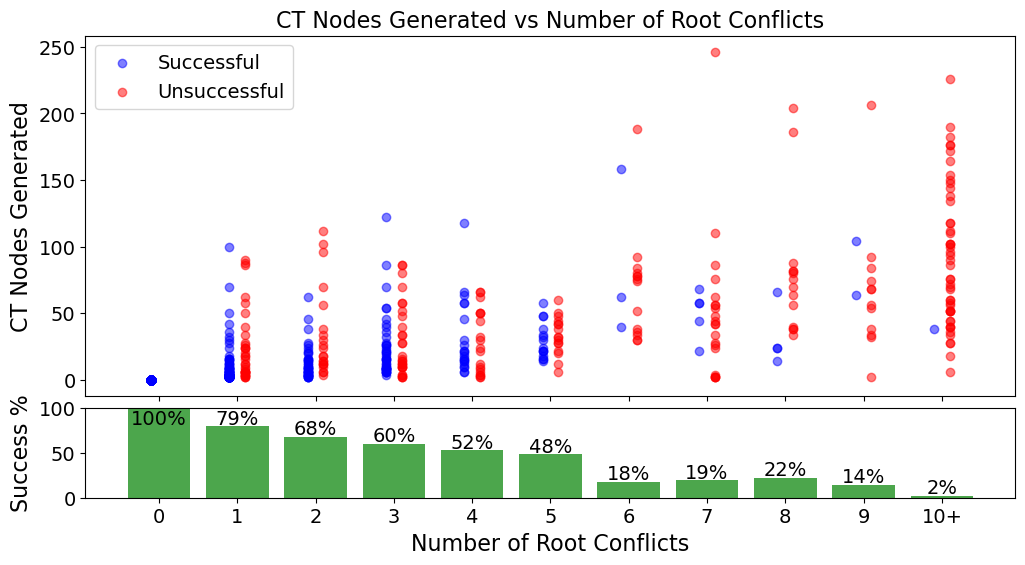}
    \caption{We show results on MAMP start-goal problems with different solvers and categorize them by the number of conflicts in the root node (a proxy for problem difficulty). The top plot shows the number of CT nodes that the CBS Protocol generates for successful (blue) and unsuccessful (red) instances while the bottom  shows the success rate.    
    }
    \label{fig:results-success-rates}
    \vspace{-1em}
\end{figure}

Fig \ref{fig:results-success-rates} shows the success rate (bottom) and number of CT Nodes generated (top) based on the number of root conflicts. The CBS Protocol solved 54\% of instances with non-zero conflicts. As the number of conflicts increases, the number of CT Nodes generated for successful instances increases (e.g., the median of the blue dots increases). We also observe that the overall success rate decreases as root conflicts increase. 

We emphasize that the CBS Protocol is tightly coupled to the success rate and speed of the underlying single-agent APIs as it repeatedly calls them during its search process. For example, the RL and Collocation solvers had relatively low success rates with constraints (35\%, 46\% respectively) compared to A* with Dubin's or 4-connected dynamics (87\%, 94\%). Additionally, some solvers/APIs took up to 10 seconds to satisfy a single query, which limited the number of CT nodes CBS could explore within the 2 minute timeout.
Collision checking also took a non-negligible amount of time for larger amounts of agents. 

We conducted an ablation study by removing ``experience" from each single-agent API and rerunning experiments. This did not noticeably change the overall CBS protocol success rate, but substantially increased the median API query time, ranging from 10\% to 300\% increases depending on the solver, except for Collocation which took 20x longer.


Importantly, the raw results should not be interpreted as a fundamental limitation of the CBS Protocol. With optimized single-agent APIs and collision checking (e.g., planning times under 0.1 seconds), CBS is capable of generating thousands of nodes and achieving substantially higher success rates. Thus, our findings highlight the dependency of CBS performance on the single-agent API and illustrate the trend of CBS's CT node generation and conflict resolution.

\textbf{Baseline: } As mentioned in Section \ref{sec:ah-mamp}, algorithmically heterogeneous baselines are non-existent. We did not implement a local replanning baseline as our individual APIs are too slow to be called often during execution (which may also be an issue in real applications). Instead, we let each agent plan individually and implemented a local priority-based collision avoidance policy. When a lower-priority agent is within a radius of 0.1 meters to a higher-priority agent, the lower-priority agent attempts to reverse direction and backtrack. If agents are within 0.05 meters, then they freeze. This baseline worked on 87\% of instances without root conflicts and 9\% of instances with root conflicts.

\textbf{Heterogeneous Tasks:}
We do a proof-of-concept and replace the start-goal task in the plan API with different tasks.
For A* and RRT we implement coverage planners that need to visit a set of points in any arbitrary order. For optimization (direct collocation), the agent attempts to maintain a radius around a given ``surveillance" point.
For diffusion, we train a policy to mimic motion patterns on data where the robot visits an intermediate point before proceeding to the goal.
Finally, for the RL solver we implement a Roomba-like policy that tries to sweep the entire environment. Due to space constraints, Figure \ref{fig:case-study} shows a single example solution with various APIs (more examples are provided on the website). We highlight that we only need to input the different plan APIs into the CBS Protocol without any additional changes.



\section{Conclusion and Future Directions}

The future of robotics will require multiple robots from different companies completing individual tasks to efficiently find collision-free paths in shared environments. It is unclear how to effectively do this due to each robot having its own interfaces and motion planning software.
To that end, we show how we can use Conflict-Based Search \cite{sharon2015cbs} as a protocol. The CBS Protocol solely requires agents to specify a single \texttt{plan()} API that takes in certain space-time constraints and returns a path with corresponding cost and space-time volume. We show how the CBS Protocol enables finding collision-free paths for a variety of solvers (A*, RRT, Direct Collocation, Diffusion, Reinforcement Learning), and is not limited to start-goal tasks but can work with arbitrary independent motion planning tasks. 

\textbf{Future Work:}
From an industry perspective, we would like to see the CBS Protocol adopted and deployed on real heterogeneous systems with efficient solvers. 

 
From the single-agent \texttt{plan()} API perspective, there is significant potential in developing better solvers that handle CBS space-time constraints and replanning. In particular, these space-time constraints are a special subcategory of ``dynamic" obstacles whose structure could be exploited for efficient planning. 
More effective versions of reusing experience in respect to these constraints is also promising.

From the AH-MAMP and multi-agent protocol perspective, the use of heterogeneous \texttt{plan()} APIs opens up interesting questions on how to best leverage their strengths and weaknesses. For example, in our experiments, different APIs had varying success rates and query times. Enabling the CBS Protocol to intelligently reason about this would likely boost performance. Additionally, as a proof of concept, the current version of the CBS Protocol retools CBS from classical MAPF but does not include additional improvements in the MAPF literature. 
Achieving this would likely require expanding the single-agent \texttt{plan()} API inputs. 

From the most broad perspective, we hope this work bridges a gap between single-agent motion planning, multi-agent motion planning, and MAPF, as well as encourages future development of multi-agent motion planning protocols that can handle heterogeneous single-agent solvers.

\section*{Author Contribution Statements}
\begin{itemize}
    \item RV conceived the project, gathered the team, implemented the CBS Protocol and parts of single-agent solvers, ran all experiments, and wrote the paper.
    \item AT developed the A* and RRT single-agent solvers, RV assisted and advised.
    \item HH and SZ developed the RL single-agent solver, RV assisted and advised.
    \item VS and YC developed the direct collocation single-agent solver, RV, GO, and JA assisted and advised.
    \item ZJ trained the diffusion solver from MMD, RV and YS assisted and advised.
    \item JA, YS, JL, ML gave feedback on the paper.
    \item JL and ML advised the overall project.
\end{itemize}

\bibliographystyle{IEEEtran}
\bibliography{references}

\clearpage
\setcounter{figure}{0}
\renewcommand{\thefigure}{A\arabic{figure}}
\setcounter{table}{0}
\renewcommand{\thetable}{A\arabic{table}}
\setcounter{section}{0}
\renewcommand{\thesection}{A\arabic{section}}

\section*{Appendix}

\section{Quick Summary}

\subsection{Motivation}

As single-robot systems continue to grow more capable, the next frontier will be environments populated by multiple autonomous agents—likely from different manufacturers—operating side by side. Imagine, for example, a factory floor where autonomous forklifts, carts, and trucks all move independently while pursuing their own objectives. In such settings, a central challenge emerges: how can we reliably compute collision-free motions for heterogeneous agents?

Crucially, each robot may rely on its own motion-planning software (possibly proprietary), embody different morphologies and kinodynamic constraints, and pursue distinct tasks. Any framework for safe coordination must therefore reason across diverse solvers, embodiments, and objectives, without assuming a shared planning infrastructure.

\subsection{Intended Takeaways}

\subsubsection{Defining a new problem setting}  
To our knowledge, the problem of \emph{multi-agent motion planning (MAMP) with algorithmically heterogeneous single-agent solvers} has not been addressed by the community. Current heterogeneous MAMP methods are not algorithmically heterogeneous as they require the different robots to all run the same planner \cite{debord2018multi_drone_ground_cbs,solis2021representation_optimal_cbs} or focus on heterogeneous behaviors (and not collision-free movement) \cite{rizk2019heterogeneous_multi_robot_survey,bettini2023heterogeneous_behavior_marl}. We introduce and formalize this new setting as \textbf{Algorithmically Heterogeneous MAMP (AH-MAMP)}. Solutions for AH-MAMP cannot rely on accessing the internals of each single-agent solver. Instead, they must be formulated as \emph{protocols} that interact with single-agent planners only through well-defined APIs.  

\subsubsection{Proposing CBS as a Protocol}  
Our second contribution is to cast Conflict-Based Search (CBS) \cite{sharon2015cbs}---a foundational method in multi-agent path finding \cite{stern2019mapfbenchmark}---as such a protocol. We observe that CBS fundamentally relies on a single-agent \texttt{plan()} interface that can return a path subject to space-time constraints. Specifically, the API requires:  
\begin{itemize}
    \item \textbf{Input:} A list of space-time constraints.  
    \item \textbf{Output:} A path satisfying these constraints, its occupied space-time volume (possibly in compact form), and the solution cost. Returns None if no solution is found.  
\end{itemize}  
Constraints minimally define space-time locations $(p,t)$ that must be avoided, but can be generalized to space-time volumes for practical performance. Crucially, CBS operates independently of how the underlying \texttt{plan()} call is implemented, enabling it to coordinate diverse single-agent solvers under a shared protocol abstraction.  

This perspective highlights a practical benefit: new single-agent methods can be easily integrated into the CBS Protocol as long as they expose the required API. For example, our work demonstrates how to use an RL policy within CBS—a ``novel'' contribution from the RL or CBS perspective but straightforward under the API abstraction.  

\subsubsection{Demonstrating broad applicability}  
We demonstrate that the CBS Protocol can coordinate a heterogeneous team of robots equipped with a wide range of solvers: heuristic search (e.g., A*), sampling-based planning (e.g., RRT), trajectory optimization (e.g., direct collocation), diffusion-based planning, and reinforcement learning. 
Additionally, we show how the CBS Protocol is task-agnostic: the \texttt{plan()} API is not limited to start-goal planning and can plan independent motion-planning tasks like coverage or surveillance.

\subsection{Main Limitations}

\subsubsection{Independent tasks}  
By design, the CBS Protocol assumes that agents solve independent tasks. It does not directly address collaborative multi-agent tasks where agents must explicitly coordinate behaviors or share objectives. In such cases, task- or robot-specific logic is required, which lies outside the scope of a task-agnostic protocol. Nonetheless, CBS can still mitigate collisions between \emph{groups} of collaborating agents by treating them as meta-agents.  

\subsubsection{Dependency on single-agent performance}  
The efficiency of the CBS Protocol is directly tied to the quality of the underlying single-agent solvers. In our experiments, some solvers exhibited low success rates or required up to 10 seconds to return a plan, which bottlenecked overall performance. This is not an inherent limitation of the CBS Protocol itself: faster and more reliable single-agent APIs would immediately improve scalability.

\subsection{Future Work}
\subsubsection{Future work for single-agent practitioners}
Developing more efficient single-agent solvers for the CBS Protocol, i.e., that handles space-time constraints and replanning, will enable more efficient multi-agent motion planning with the CBS Protocol. The benefit of the CBS Protocol is that these single-agent solvers can be developed independently and still be used plug-and-play with the CBS Protocol.

\subsubsection{Future work for multi-agent practitioners}
As a proof-of-concept, the CBS Protocol took the core CBS idea without leveraging modern advancements in classical CBS. Thus, promising future work would be to incorporate improvements like stronger constraints and Enhanced CBS \cite{barer2014suboptimalecbs}. Interleaving partial planning and execution \cite{rhcrLi2020} could also replace the need for a single long planning time in the beginning with interspersed planning throughout. Achieving this would likely require expanding the single-agent \texttt{plan()} API inputs. 

More broadly, we hope more attention will be given to the AH-MAMP problem setting which will become prevalent as single-agent robots mature. We hope to see more solutions/protocols in the future.

\section{Additional Experimental Details \& Results}

\begin{table*}[htbp]
\centering
\begin{tabular}{|c|c|c|c|c||c|c|c|}
\hline
\multirow{2}{*}{Solver} & \multirow{2}{*}{Dynamics} & \multirow{2}{*}{Shape (Size)} & \multirow{2}{*}{Maps} & \multirow{2}{*}{Timeout (s)} & \multicolumn{3}{c|}{No experience / Experience} \\ \hhline{|~|~|~|~|~||-|-|-|}
 & & & & & Mean Times (s) & Median Times (s) & Success \% \\ \hhline{|=|=|=|=|=#=|=|=|}
\multirow{3}{*}{A*} & Ackermann & \multirow{2}{*}{Rectangle (0.4x0.1)} & \multirow{3}{*}{All} & \multirow{3}{*}{10} & 5.62 / 4.74 & 4.60 / 2.09 & 57.1 / 56.4 \\
 & Dubins &  &  &  & 2.09 / 1.85 & 0.26 / 0.17 & 88.2 / 87.7 \\ \hhline{|~|-|-|~|~||~|~|~|}
 & 4-connected & Rectangle (0.8x0.3) &  &  & 1.41 / 1.58 & 0.72 / 0.66 & 91.3 / 94.0 \\ \hline
Colloc & Ackermann & Rectangle (0.8x0.3) & empty, random & 10 & 6.45 / 0.67 & 9.42 / 0.44 & 44.9 / 46.5 \\ \hline
Diffusion & None & Circle (0.15 radius) & Ribera & 10 & 1.96 / 0.68 & 1.59 / 0.51 & 61.9 / 62.3 \\ \hline
RL & Unicycle & Oval-like (0.8x0.3) & All$^\dagger$ & 5 & 1.67 / 1.20 & 1.72 / 0.72 & 34.4 / 35.0 \\ \hline
\multirow{2}{*}{RRT} & 4-connected & \multirow{2}{*}{Triangle (0.6x0.4)} & \multirow{2}{*}{All} & \multirow{2}{*}{5} & 1.38 / 1.28 & 0.44 / 0.17 & 79.1 / 82.3 \\
 & None &  &  &  & 0.87 / 0.82 & 0.18 / 0.07 & 86.1 / 82.9 \\ \hline
\end{tabular}
\caption{Additional information on single agent start-goal solvers. The A* and RRT solvers worked reliably across all maps. Direction Collocation (Colloc) took too long on the IN2D maps, Diffusion struggled with small obstacles or large maps, and $^\dagger$ denotes how the RL agent was unreliable on IN2D maps. The timeout denotes the single-agent \texttt{plan()} time limit before returning failure. The statistics on the right columns report statistics for replanning without and with experience in the CBS Protocol (e.g., the first results 5.62 / 4.74 denote that without experience the mean runtime is 5.62 seconds, while with experience it is 4.74). We see that experience noticeably decreases the median runtime across all solvers.}
\label{tab:app-sas-results}
\end{table*}

\subsubsection{Start-Goal Solvers} Table \ref{tab:app-sas-results} describes additional information about the single agent start-goal solvers. The success percentage (last column) is the success rate of the single-agent \texttt{plan()} returning a solution that satisfies constraints within the timeout. Mean and median times denote the \texttt{plan()} time for re-planning given constraints, with failures included as the timeout.

\subsubsection{Non-Start Goal Solvers} We describe in more detail the heterogeneous tasks described in Section \ref{sec:main-results}. 
For tasks which are very-long or infinite horizon (e.g., coverage which could continue forever), we had the \texttt{plan()} APIs return a path of at most 50 simulation timesteps.

For the coverage A* and RRT \texttt{plan()} API, we sampled 4 random states from the map and have the planners return a path that visits all 4 states (or a partial path that reaches the 50 simulation timestep limit). This is done by iteratively planning paths to the nearest state. 

For the Direction Collocation ``surveillance" task, we change the objective function so that it penalizes deviations from a target radius distance from a target point and forward velocity. Solving this induces a circular motion around the target point.

For the RL coverage planner, we discretized the environment into cells and greedily set the goal for the RL agent to reach the next location it has not covered yet.

For the Diffusion solver, we collected a dataset of start-goal paths where each path visits the same central location on the way to its goal. We can then train and evaluate the model identical to that for a normal start-goal dataset, except that now the model will exhibit the new motion patterns in the dataset.

\subsection{Visualizations}
\begin{figure*}[htbp]
  \centering
  
  \begin{subfigure}[b]{0.48\textwidth}
    \centering
    \includegraphics[width=\linewidth]{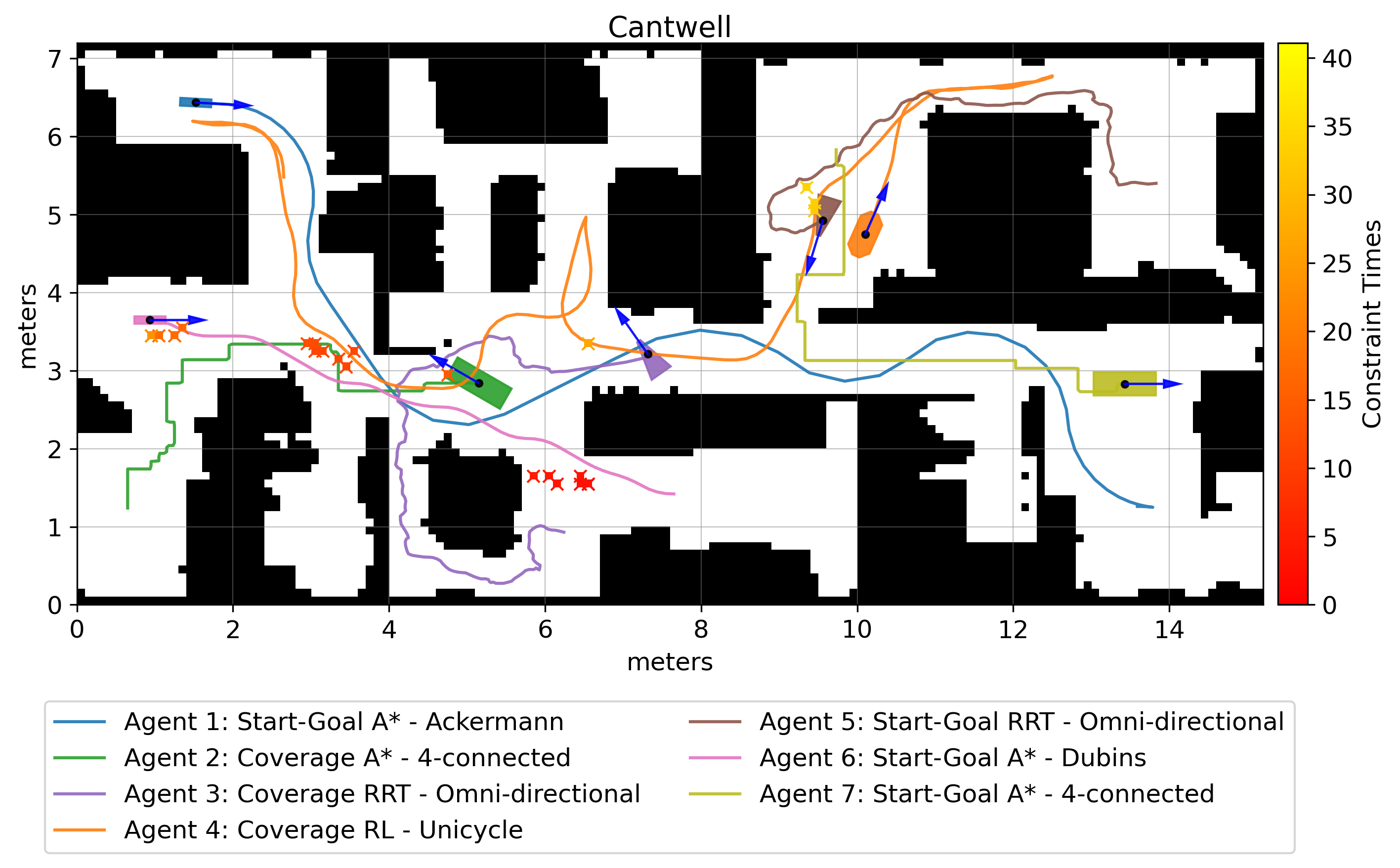}
    \label{fig:in2d-cantwell-1}
  \end{subfigure}\hfill
  \begin{subfigure}[b]{0.48\textwidth}
    \centering
    \includegraphics[width=\linewidth]{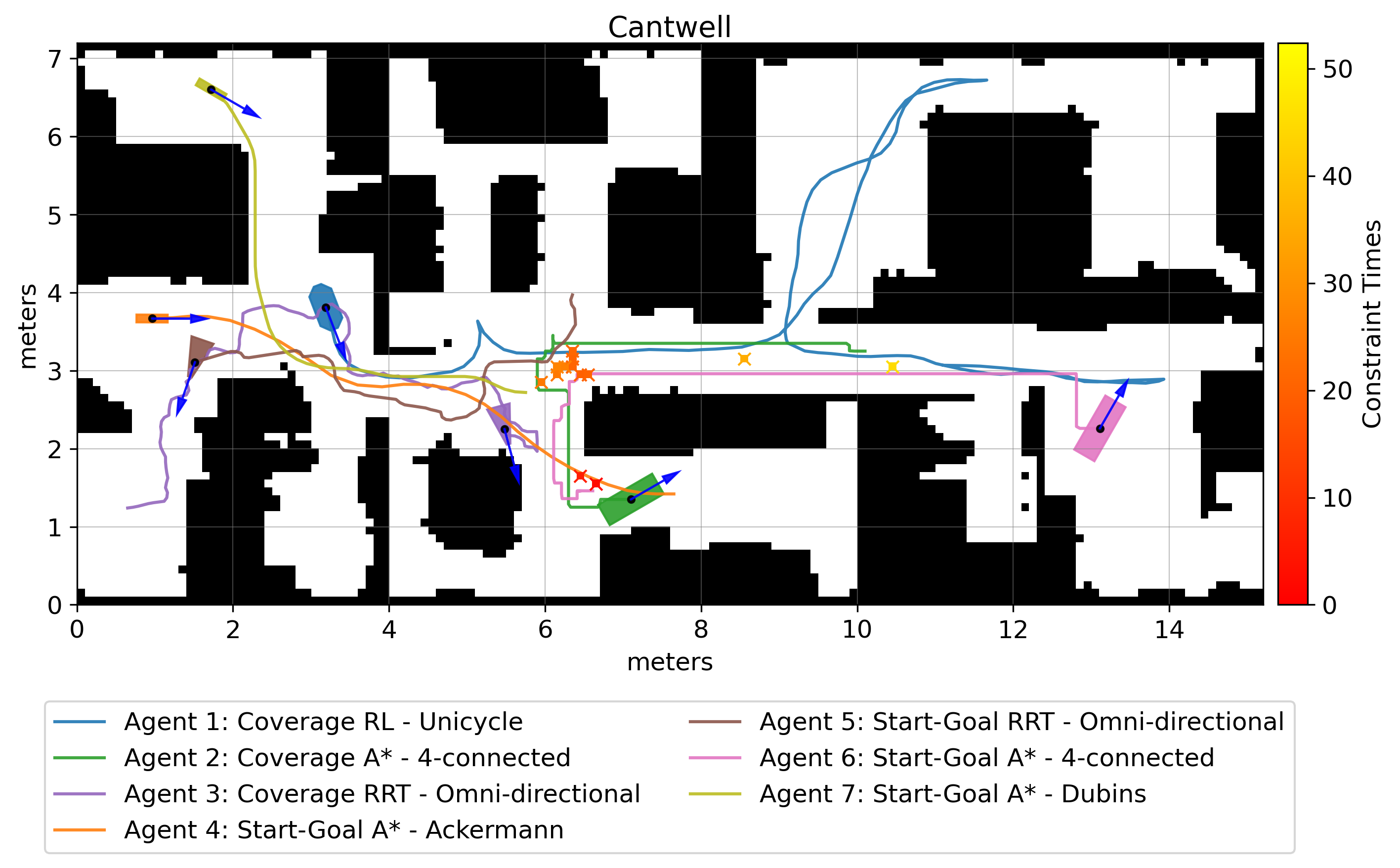}
    \label{fig:in2d-cantwell-2}
  \end{subfigure}

  \vspace{0.2em} 

  \begin{subfigure}[b]{0.48\textwidth}
    \centering
    \includegraphics[width=\linewidth]{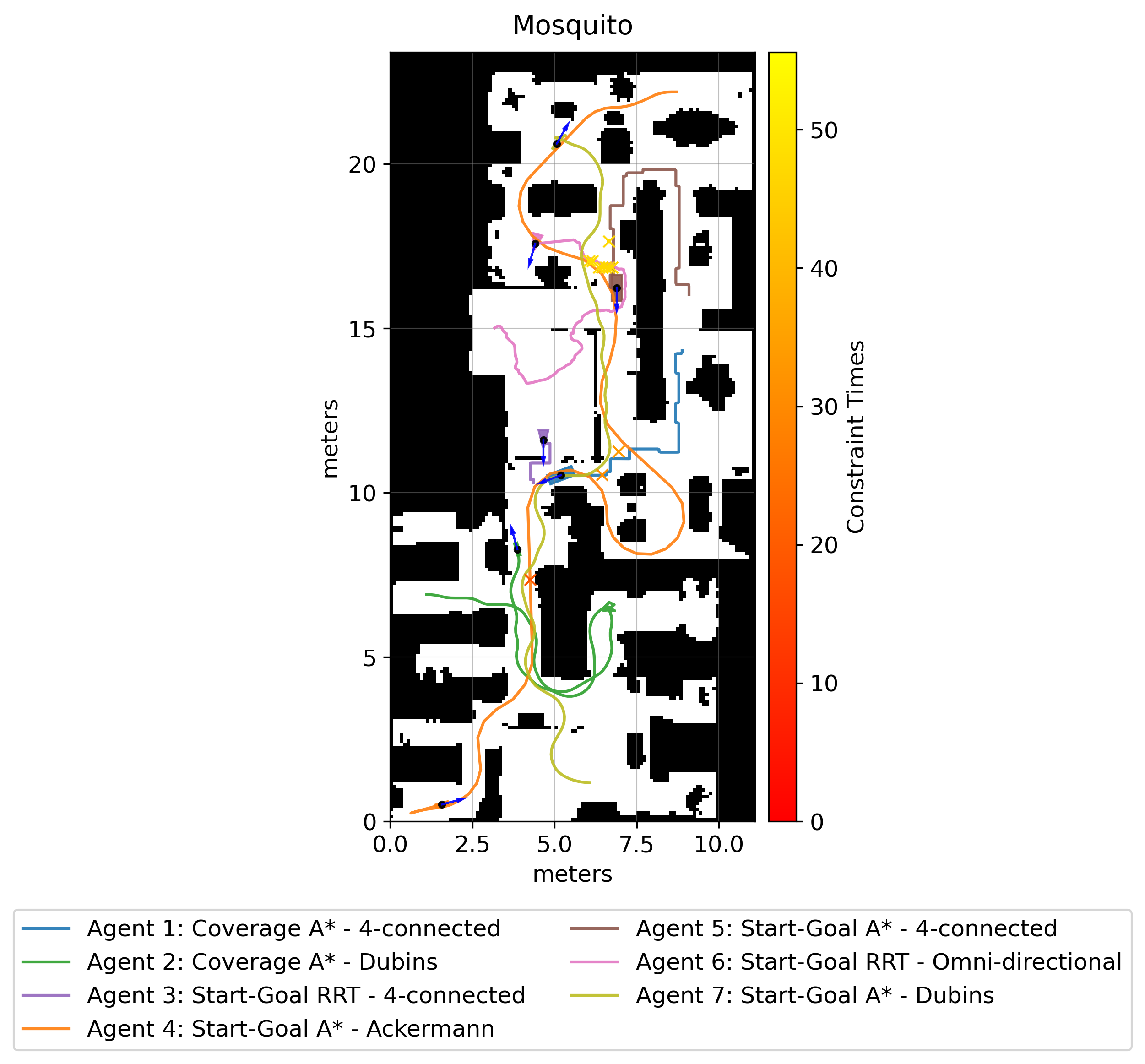}
    \label{fig:in2d-mosquito-1}
  \end{subfigure}\hfill
  \begin{subfigure}[b]{0.48\textwidth}
    \centering
    \includegraphics[width=\linewidth]{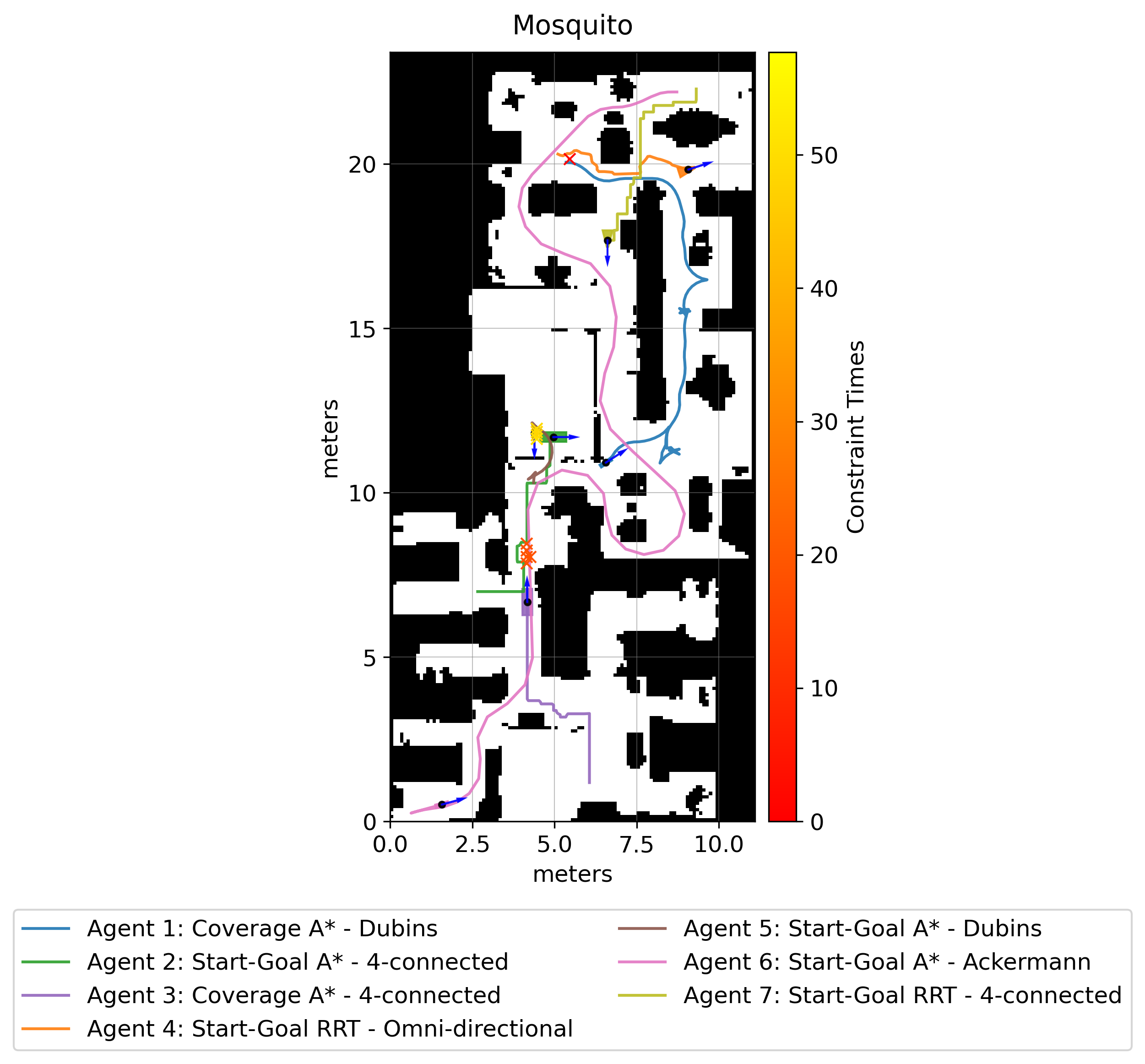}
    \label{fig:in2d-mosquito-2}
  \end{subfigure}

  \vspace{0.2em}

  \begin{subfigure}[b]{0.48\textwidth}
    \centering
    \includegraphics[width=\linewidth]{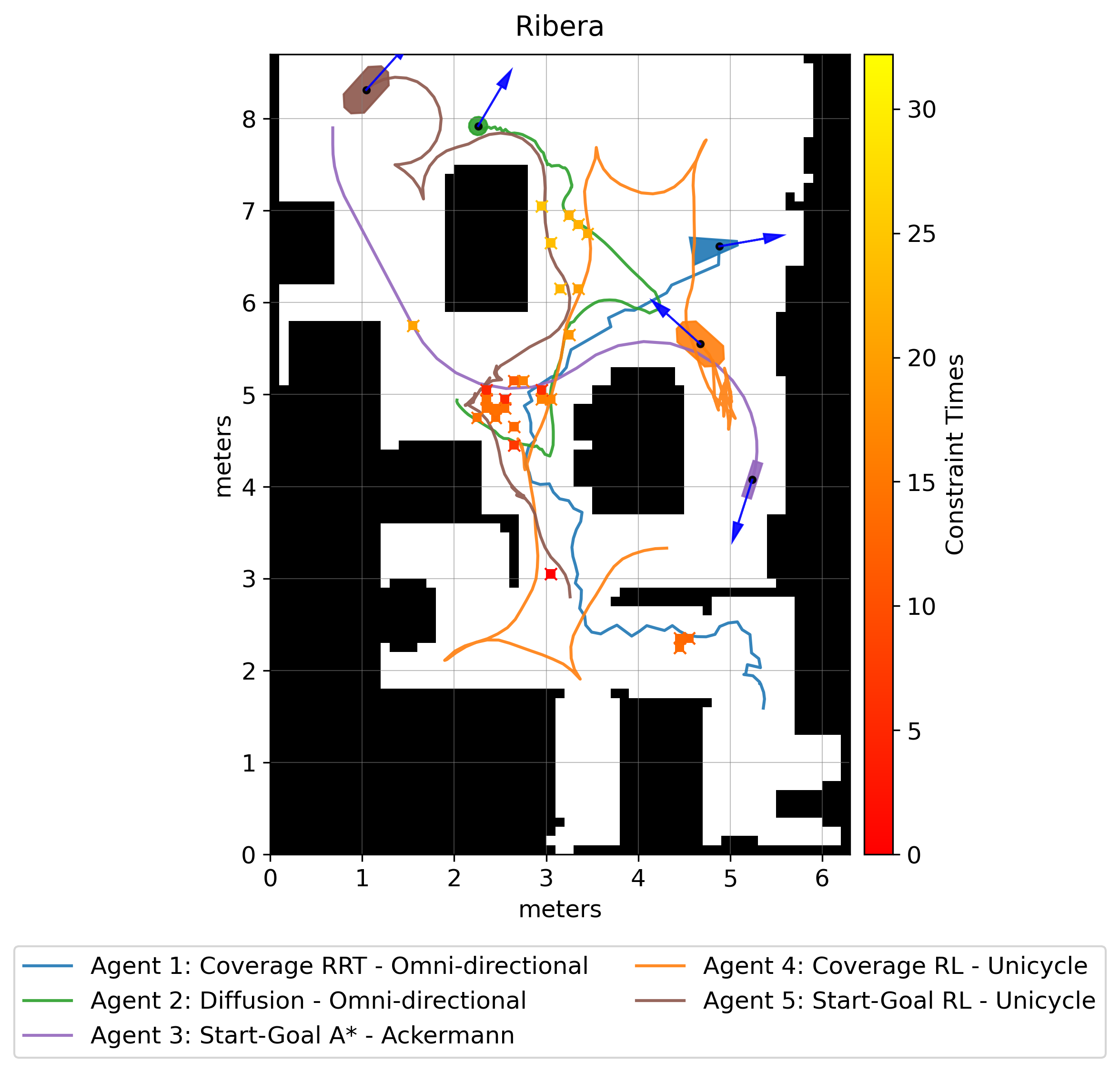}
    \label{fig:in2d-ribera-1}
  \end{subfigure}\hfill
  \begin{subfigure}[b]{0.48\textwidth}
    \centering
    \includegraphics[width=\linewidth]{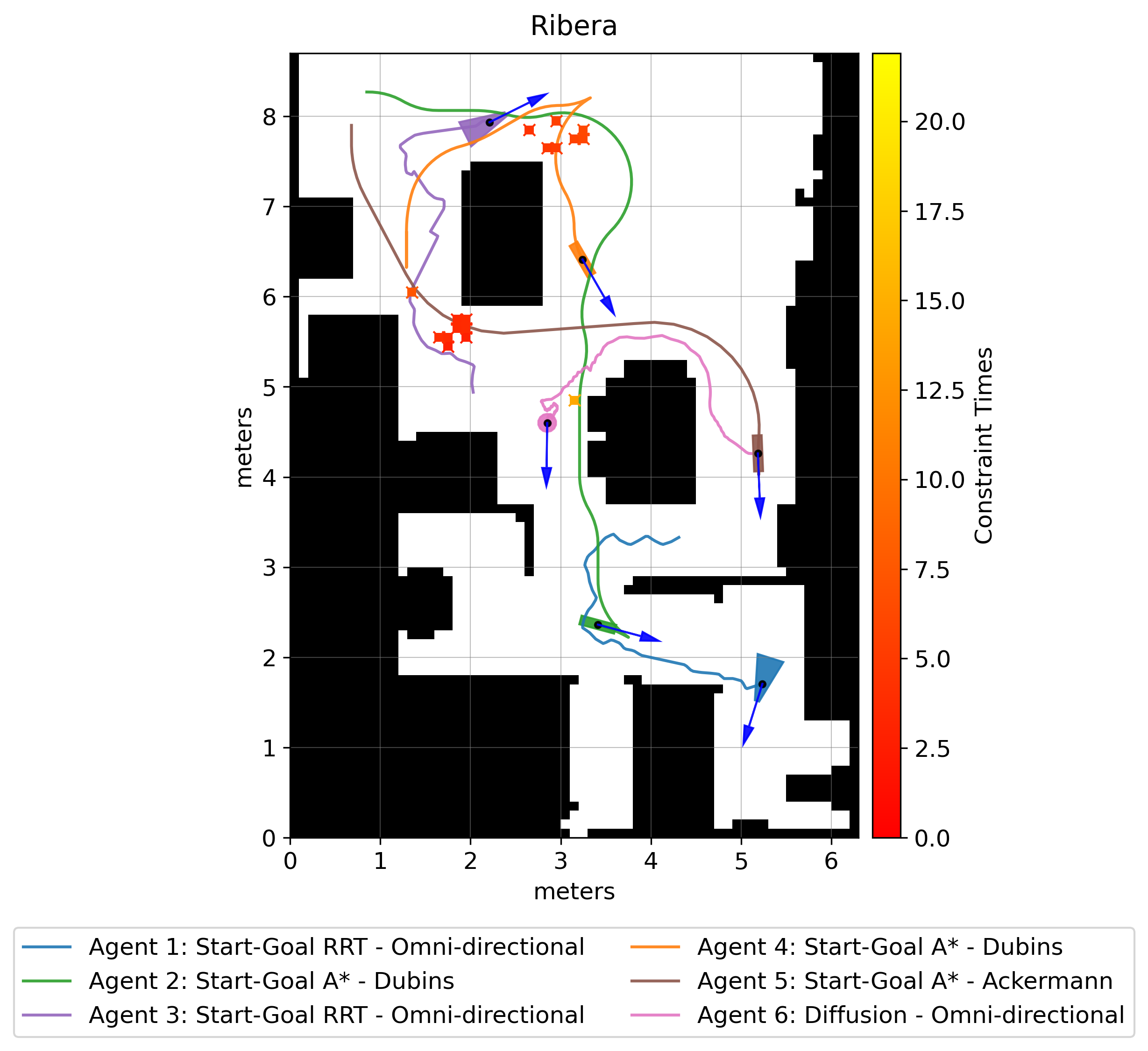}
    \label{fig:in2d-ribera-2}
  \end{subfigure}

  \caption{Examples of successful instances of the CBS Protocol with algorithmically heterogeneous agents on IND2 Maps. Each legend denotes the task, solver, and dynamics for each agent. Note that some agents also have different footprint shapes/sizes. X's denote space-time constraints required for the solution found by CBS, with the time depicted by the color bar.}
  \label{fig:in2d-overall}
\end{figure*}

\begin{figure*}[htbp]
  \centering
  
  \begin{subfigure}[b]{0.48\textwidth}
    \centering
    \includegraphics[width=\linewidth]{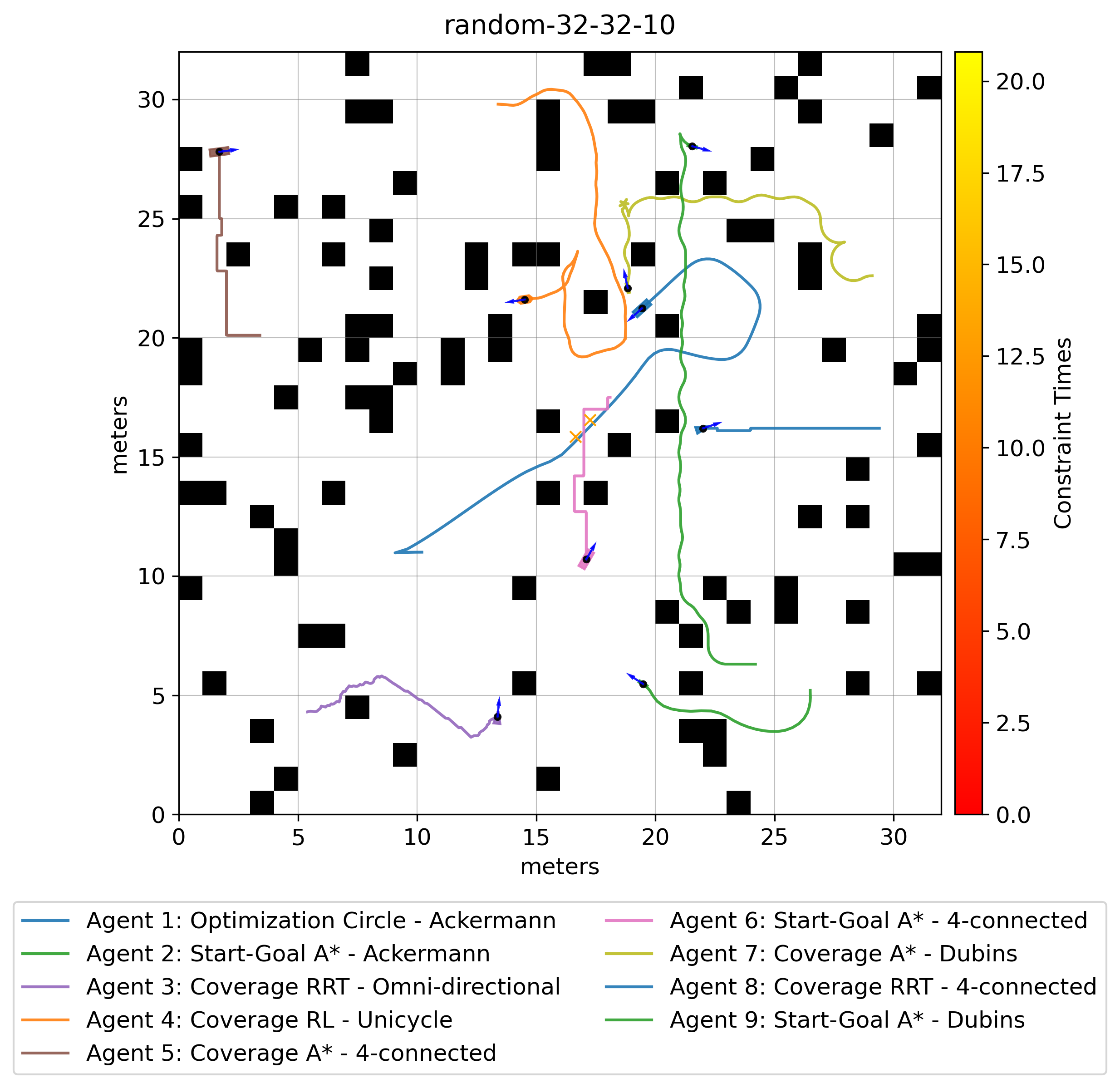}
    \label{fig:sub7}
  \end{subfigure}\hfill
  \begin{subfigure}[b]{0.48\textwidth}
    \centering
    \includegraphics[width=\linewidth]{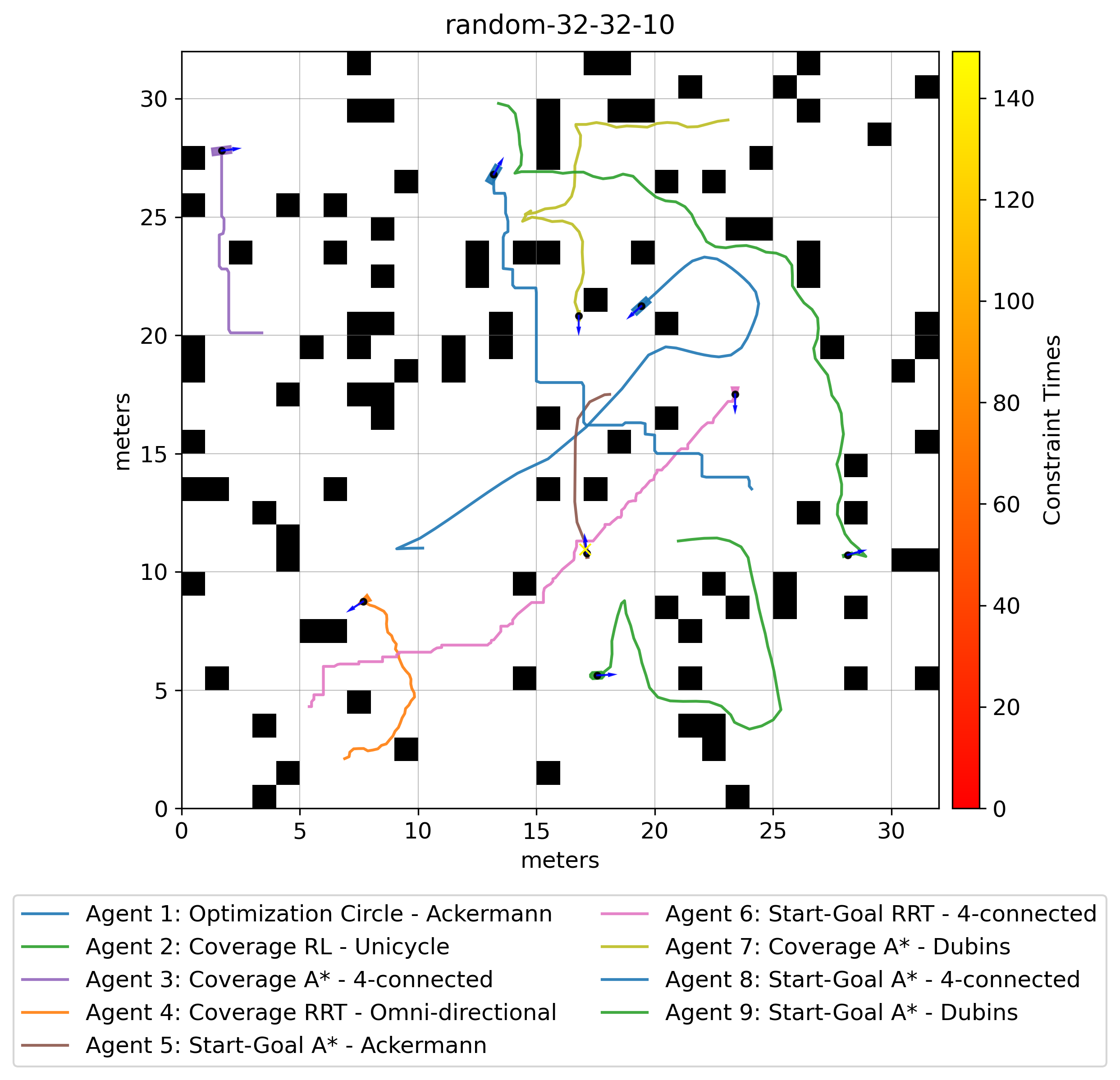}
    \label{fig:sub8}
  \end{subfigure}
  
  \vspace{0.8em}
    
  \begin{subfigure}[b]{0.48\textwidth}
    \centering
    \includegraphics[width=\linewidth]{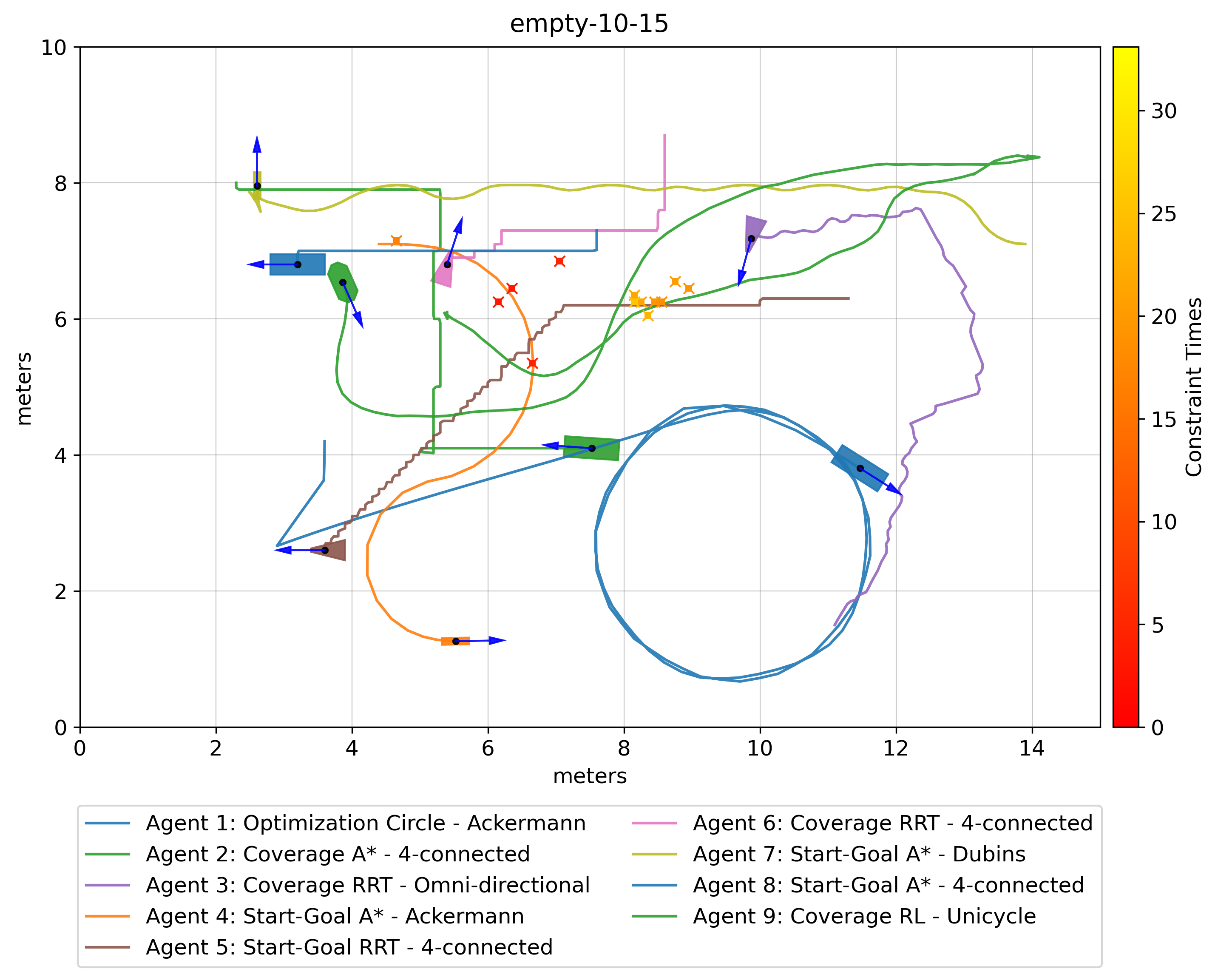}
    \label{fig:sub3}
  \end{subfigure}\hfill
  \begin{subfigure}[b]{0.48\textwidth}
    \centering
    \includegraphics[width=\linewidth]{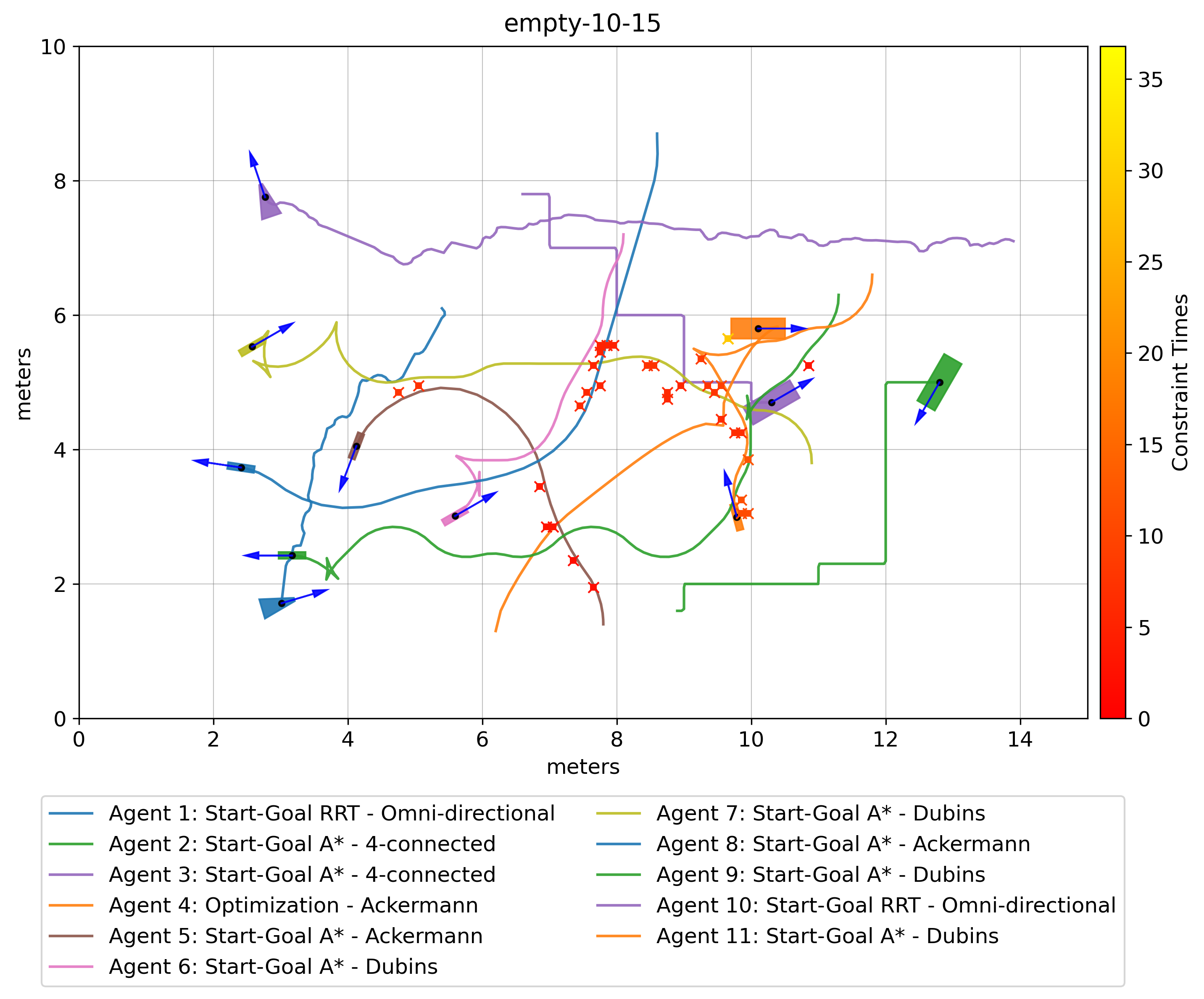}
    \label{fig:sub4}
  \end{subfigure}

  \caption{Examples of successful instances of the CBS Protocol with algorithmically heterogeneous agents on a map with random obstacles and an empty map. Each legend denotes the task, solver, and dynamics for each agent. Note that some agents also have different footprint shapes/sizes. X's denote space-time constraints required for the solution found by CBS, with the time depicted by the color bar.}
  \label{fig:empty-random-overall}
\end{figure*}

Figures \ref{fig:in2d-overall} and \ref{fig:empty-random-overall} visualize results for heterogeneous tasks which were solved using the CBS Protocol. We increased the timeouts of the Direct Collocation solver to 30 seconds so that it succeeded on more instances, and increased the overall CBS Protocol search time to 300 seconds. Instances were generated by randomly sampling single agent solvers and tasks.

\end{document}